\documentstyle[amsfonts,11pt]{article}

\def\ber#1#2{\begin{equation}\begin{array}{#1}\displaystyle{#2}}
\def\ber#1{\begin{equation}\begin{array}{#1}\displaystyle}
\def\bernn#1#2{$$\begin{array}{#1}\displaystyle{#2}}

\def\eer#1{\end{array}\label{#1}\end{equation}}
\def\eernn{\end{array}$$}
\def\r#1#2{\noindent\hbox{\hbox to 24 pt{\hfil[#1]~}%
\vtop{\hsize = 12.5 truecm\noindent#2}}\vskip 5 pt\vfil}
\def\chap#1#2#3{\noindent\hbox{\hbox to 1.5 truecm{\hfil#1}%
\hbox to 14 truecm{~#2\leaders\hbox to 0.5 em{\hfil.\hfil}\hfill#3}}\par}
\def\cchap#1#2#3#4{\noindent\hbox{\hbox to 1.5 truecm{\hfil#1}%
\hbox to 14 truecm{~#2\hfil}}\par
\noindent\hbox{\hskip 1.5 truecm%
\hbox to 14 truecm{~#3\leaders\hbox to 0.5 em{\hfil.\hfil}\hfill#4}}\par}
\def\bbt{\bibitem}
\def\be{\begin{equation}}
\def\en{\end{equation}}
\def\ber{\begin{eqnarray}}
\def\enr{\end{eqnarray}}
\def\nmb{ \nonumber\\}
\def\d{\partial}
\def\rbr{\rbrack}
\def\lbr{\lbrack}
\def\rbrc{\rbrace}
\def\lbrc{\lbrace}
\def\ov{\over }
\def\tld{\tilde}
\def\brv{\breve}

\def\eq{\equiv}

\def\MTR{Manin triple }

\def\DLG{double Lie group }
\def\Tta{\Theta}
\def\sgm{\sigma}
\def\Sgm{\Sigma}

\def\im{\imath}
\def\rh{\rho}

\def\Lm{\Lambda}
\def\Om{\Omega}

\def\et{\eta}
\def\eps{\epsilon}
\def\dlt{\delta}
\def\Gm{\Gamma}
\def\Sgm{\Sigma}

\begin{document}
\rightline{Landau Tmp/09/97.}
\rightline{September 1997}
\vskip 2 true cm
\centerline{\bf MIRROR SYMMETRY AS A POISSON-LIE T-DUALITY.}
\vskip 2.5 true cm
\centerline{\bf S. E. Parkhomenko}
\centerline{Landau Institute for Theoretical Physics}
\centerline{142432 Chernogolovka,Russia}
\vskip 0.5 true cm
\centerline{spark@itp.ac.ru}
\vskip 1 true cm
\centerline{\bf Abstract}
\vskip 0.5 true cm

 The transformation properties of the $N=2$ Virasoro superalgebra
generators under Poisson-Lie T-duality in (2,2)-superconformal
WZNW and Kazama-Suzuki models is considered.
It is shown that Poisson-Lie T-duality
acts on the $N=2$ super-Virasoro algebra generators
as a mirror symmetry does: it unchanges the generators from one
of the chirality sectors while in another chirality sector     
it changes the sign of $U(1)$ current and interchanges spin-3/2 
currents. We discuss Kazama-Suzuki models generalization of this
transformation and show that Poisson-Lie T-duality acts
as a mirror symmetry also.



\smallskip
\vskip 10pt
\centerline{\bf Introduction.}

The $N=2$ superconformal field theories (SCFT's) play an important role
in diverse aspects of superstring theory.
As the first application in the superstrings the $N=2$ SCFT's 
was appeared in Gepner's superstring vacua construction
~\cite{Gep} (see also ~\cite{Tao}), where it was shown
that the $N=2$ SCFT's may describe Calabi-Yau manifolds compactifications
of the superstrings. Since then the $N=2$ SCFT's, and the profound structures
assotiated with them are an area of investigation.
On the other hand, it is well known that the
string vacua, considered as conformal field thoeries, in general,
have deformations under which the geometry of the target
space changes. These deformations include the discrete duality
transformations, so called T-duality, which are symmetries of the
underlying conformal field theory ~\cite{GiPR}, ~\cite{AlvG}.

 Mirror symmetry ~\cite{MS} discovered in superstring theory
is the special type of T-duality. At the level of conformal
field theory, it could be formulated as an isomorphism between
two theories, amounting to a change of sign of the $U(1)$
generator and interchange spin-3/2 generators of the left-
moving (or right-moving) $N=2$ superconformal algebra.

 Mirror symmetry has mostly been studied in the context of
Calabi-Yau superstring compactifications. Though this approach 
is quite general, since it allows for the moduli of the theory
to be varied, its quantization seems to be fuzzy
because it is not known what is the quantum Calabi-Yau $\sgm$-model.
In fact the only rigorously established example of mirror symmetry,
the Green-Plesser construction ~\cite{GrPles}, is based on the tensor
products of the $N=2$ minimal models ~\cite{MM}. 

 In this note we propose algebraic approach to the mirror symmetry
based on Poisson-Lie (PL) T-duality. The Poisson-Lie T-duality, 
recently discoverd by C. Klimcik and P. Severa  ~\cite{KlimS1} is a 
generalization of the standard non-Abelian T-duality 
~\cite{OsQ}-~\cite{CurZ}.
This generalized duality is associated with two groups
forming a Drinfeld double ~\cite{Drinf1} and the duality transformation
exchanges their roles. This approach has recieved futher developments
in the series of works ~\cite{KlimS2}-~\cite{TyU}.

 The supersymmetric generalization
of PL T-duality was considered in~\cite{Sfet}-~\cite{P2}.
In particular, in ~\cite{P2} it was shown that PL T-duality
in $N=2$ SWZNW models is governed by the complex Heisenberg doubles
associated with the group manifolds of the models and PL T-duality
mappings are given on-shell by the special super Kac-Moody gauge
transformations thus establishing (on-shell) PL self-duality of the 
$N=2$ SWZNW models.  

 In the recent paper of Klimcik ~\cite{SuperKl} (1,1) supersymmetric
PL T-duality was formulated off-shell. It proves PL T-duality
in $N=2$ SWZNW models.

 The present note is based on the results of ~\cite{P2}
(so it can be considered as its appendix) and  devoted to the 
investigation (on-shell) of the transformation properties of the 
$N=2$ super-Virasoro algebra under the PL T-duality in the $N=2$ SWZNW 
and Kazama-Suzuki models ~\cite{KaSu}.
We will show that PL T-duality transforms the generators of the $N=2$ super-
Virasoro algebra exactly the same way as the mirror duality does. 
Thus we
obtain more rigorous arguments for the conjecture proposed in ~\cite{GiWi}
that mirror symmetry can be related to a gauge symmetry of the
self-dual points of the $N=2$ SCFT's moduli space (the N=0 version
of this conjecture see in ~\cite{GiKir}).

 In the section 1 we briefly review the results of paper ~\cite{P2}.
In section 2 we describe transformation properties of N=1 super-Kac-Moody
algebra currents under PL T-duality. Then we describe in section 3
PL T-duality transformations of the left-moving and right-moving
$N=2$ super-Virasoro algebras in the $N=2$ SWZNW and Kazama-Suzuki models.

\vskip 10pt
\centerline{\bf1. The classical $N=2$ superconformal WZNW model.}

 In this section we briefly review the $N=2$ SWZNW models
following  ~\cite{swzw, QFR2, P2}.

 We parametrize super world-sheet introducing the light cone
coordinates
$x_{\pm}$, and grassman coordinates $\Tta_{\pm}$ (we shall use N=1
superfield formalism).
The generators of supersymmetry and covariant
derivatives are
\be
Q_{\mp}= {\d \ov \d\Tta_{\pm}}+\im \Tta_{\pm}\d_{\mp},\
D_{\mp}= {\d \ov \d\Tta_{\pm}}-\im \Tta_{\pm}\d_{\mp}.
\label{1}
\en
They satisfy the relations
\be
\lbrc D_{\pm},D_{\pm}\rbrc= -\lbrc Q_{\pm},Q_{\pm}\rbrc= -\im 2\d_{\pm},\
\lbrc D_{\pm},D_{\mp}\rbrc= \lbrc Q_{\pm},Q_{\mp}\rbrc=
\lbrc Q,D\rbrc= 0,
\label{2}
\en
where the brackets $\lbrc,\rbrc$ denote the anticommutator.
The superfield of N=1 SWZNW model
\be
G= g+ \im \Tta_{-}\psi_{+}+ \im \Tta_{+}\psi_{-}+
   \im \Tta_{-}\Tta_{+}F  \label{3}
\en
takes values in a real Lie group ${\bf G}$.
We will assume that its Lie algebra ${\bf g}$
is endowed with ad-invariant nondegenerate inner
product $<,>$.

The inverse group element $G^{-1}$ is defined from the relation
\be
 G^{-1}G=1 \label{4}
\en
and has the decomposition
\be
 G^{-1}= g^{-1}- \im \Tta_{-}g^{-1}\psi_{+}g^{-1}-
         \im \Tta_{+}g^{-1}\psi_{-}g^{-1}-
         \im \Tta_{-}\Tta_{+}g^{-1}(F+\psi_{-}g^{-1}\psi_{+}-
         \psi_{+}g^{-1}\psi_{-})g^{-1} \label{5}
\en

 The action of N=1 SWZNW model is given by
\ber
S_{swz}= \int d^{2}x d^{2} \Tta(<G^{-1}D_{+}G,G^{-1}D_{-}G>)   \nmb
         -\int d^{2}x d^{2}\Tta dt
          <G^{-1}\frac{\d G}{\d t},\lbrc G^{-1}D_{-}G,G^{-1}D_{+}G\rbrc>.
\label{7}
\enr
The action (\ref{7}) is invariant under the super-Kac-Moody
\ber
\dlt_{a_{+}}G(x_{+},x_{-},\Tta_{+},\Tta_{-})=
a_{+}(x_{-},\Tta_{+})G(x_{+},x_{-},\Tta_{+},\Tta_{-}), \nmb
\dlt_{a_{-}}G(x_{+},x_{-},\Tta_{+},\Tta_{-})=
-G(x_{+},x_{-},\Tta_{+},\Tta_{-})a_{-}(x_{+},\Tta_{-}),
\label{km}
\enr
where $a_{\pm}$ are ${\bf g}$-valued superfields
and N=1 supersymmetry
transformations  ~\cite{swzw}
\ber
G^{-1}\dlt_{\eps_{+}}G=(G^{-1}\eps_{+}Q_{+}G), \nmb
\dlt_{\eps_{-}}GG^{-1}=\eps_{-}Q_{-}GG^{-1}.
\label{su}
\enr

 The action of $N=2$ SWZNW model is given by (\ref{7}) also.
An additional ingredient demanded by the $N=2$ Virasoro superalgebra
of symmetries is a complex structure
on the finite-dimensional Lie algebra of the model which is
skew-symmetric with respect to the inner product $<,>$
~\cite{QFR3,QFR,QFR2}. By the definition $J$ is a complex structure
on the Lie algebra ${\bf g}$ if it is a complex structure on the
vector space ${\bf g}$ which satisfies the equation
\be
\lbr Jx,Jy \rbr-J\lbr Jx,y \rbr-J\lbr x,Jy \rbr=\lbr x,y \rbr \label{12}
\en
for any elements $x, y$ from ${\bf g}$.
It is clear that the corresponding Lie group is
a complex manifold with left (or right) invariant complex structure.
In the following we shall denote the real Lie group
and the real Lie algebra with the complex structure satisfying (\ref{12})
as the pairs $({\bf G}, J)$ and $({\bf g}, J)$ correspondingly.

 If the complex structure $J$ on the Lie algebra is fixed then it defines the
second supersymmetry transformation ~\cite{QFR2}
\ber
(G^{-1}\dlt_{\et_{+}}G)^{a}=\et_{+}(J_{l})^{a}_{b}(G^{-1}D_{+}G)^{b}, \nmb
(\dlt_{\et_{-}}GG^{-1})^{a}=\et_{-}(J_{r})^{a}_{b}(D_{-}GG^{-1})^{b},
\label{Jsu}
\enr
where $J_{l}, J_{r}$ are the left invariant and right invariant
complex structures on ${\bf G}$ which correspond to the
complex structure $J$.

 The Sugawara construction of the $N=2$ Virasoro superalgebra
generators was given in ~\cite{QFR3,QFR,QFR2,GETZ}.

 The notion of Manin triple closely related with the complex structure
on the Lie algebra. By the definition ~\cite{Drinf1},
a \MTR $({\bf g},{\bf g_{+}},{\bf g_{-}})$
consists of a Lie algebra ${\bf g}$, with nondegenerate invariant inner
product $<,>$ and isotropic Lie subalgebras ${\bf g_{\pm}}$ such that
${\bf g}={\bf g_{+}}\oplus {\bf g_{-}}$ as a vector space.

 Suppose the existence of the nondegenerate invariant inner product
$<,>$ on $({\bf g}, J)$ so that the complex structure $J$ is skew-symmetric
with respect to $<,>$ and consider the complexification
${\bf g^{\Bbb C}}$ of ${\bf g}$. Let ${\bf g_{\pm}}$ be $\pm \imath$
eigenspaces of $J$ in the algebra ${\bf g^{\Bbb C}}$
then $({\bf g^{\Bbb C}},{\bf g_{+}},{\bf g_{-}})$ is a complex \MTR.
Moreover it can be proved  that there exists the one-to-one
correspondence between the complex Manin triple endowed with antilinear
involution which conjugates isotropic subalgebras
$\tau: {\bf g_{\pm}}\to
{\bf g_{\mp}}$ and the real Lie
algebra endowed with $ad$-invariant nondegenerate inner product $<,>$
and the complex structure $J$ which is skew-symmetric with respect
to $<,>$ ~\cite{QFR3}.
The conjugation can be used to extract the real form from the complex
Manin triple.

 In this note we concentrate on the $N=2$ SWZNW models
on the compact groups (the extension on the noncompact groups is
straightforward)
that is we shall consider complex Manin triples
such that the corresponding antilinaer involutions will
coincide with the hermitian conjugations. Hence it will be implied
in the following ${\bf G}$ is a subgroup in the group of
unitary matrices and the matrix elements of
the superfield $G$ satisfy the relations:
\be
\bar{g}^{mn}=(g^{-1})^{nm},\
\bar{\psi}^{mn}_{\pm}= (\psi^{-1})^{nm}_{\pm},\
\bar{F}^{mn}= (F^{-1})^{nm}, \label{6.u}
\en
where we have used the following notations
\be
\psi^{-1}_{\pm}= -g^{-1}\psi_{\pm}g^{-1},\
F^{-1}= -g^{-1}(F+\psi_{-}g^{-1}\psi_{+}-
         \psi_{+}g^{-1}\psi_{-})g^{-1}. \label{6.not}
\en

 Now we have to consider some geometric properties of the $N=2$ SWZNW
models closely related with the existence of the complex structures
on the groups.

 Let's fix some compact Lie group with the left invariant complex
structure $({\bf G}, J)$ and consider its Lie algebra with
the complex structure $({\bf g}, J)$.
The complexification ${\bf g^{\Bbb C}}$ of ${\bf g}$ has the Manin triple
structure $({\bf g^{\Bbb C}},{\bf g_{+}},{\bf g_{-}})$. The Lie group version
of this triple  is the \DLG $({\bf G^{\Bbb C}},{\bf G_{+}},{\bf G_{-}})$
~\cite{SemTian,AlMal,LuW}, where the exponential subgroups
${\bf G_{\pm}}$ correspond to the Lie algebras
${\bf g_{\pm}}$. The real Lie group ${\bf G}$ is extracted
from its complexification with help of the hermitian conjugation $\tau$
\be
{\bf G}= \lbrc g\in {\bf G^{\Bbb C}}|\tau (g)=g^{-1}\rbrc       \label{rf}
\en
 Each element $g\in {\bf G^{\Bbb C}}$ from the
vicinity ${\bf G_{1}}$ of the unit element from ${\bf G^{\Bbb C}}$
admits two decompositions
\be
g= g_{+}g^{-1}_{-}= {\tld g}_{-}{\tld g}^{-1}_{+},  \label{13}
\en
where ${\tld g}_{\pm}$ are dressing transformed
elements of $g_{\pm}$ ~\cite{LuW}:
\be
{\tld g}_{\pm}=(g^{-1}_{\pm})^{g_{\mp}}         \label{13not}
\en
Taking into account (\ref{rf}) and (\ref{13}) we conclude that the
element $g$ ($g\in {\bf G_{1}}$) belongs to ${\bf G}$ iff
\be
\tau (g_{\pm})= {\tld g}^{-1}_{\mp}      \label{13u}
\en
These equations mean that we can parametrize the elements from
\be
{\bf C_{1}}\equiv {\bf G_{1}}\cap {\bf G} \label{13cl}
\en
by the elements from the complex group ${\bf G}_{+}$ (or ${\bf G}_{-}$),
i.e. we can introduce complex coordinates (they are just matrix elements
of $g_{+}$ (or $g_{-}$)) in the strat ${\bf C_{1}}$.
To do it one needs to solve with respect to $g_{-}$
the equation:
\be
\tau (g_{-})= (g_{+})^{g^{-1}_{-}}              \label{13c+}
\en
(to introduce ${\bf G_{-}}$-coordinates on ${\bf G_{1}}$ one needs to solve
with respect to $g_{+}$ the equation
\be
\tau (g_{+})= (g_{-})^{g^{-1}_{+}} ).            \label{13c-}
\en
Thus the formulas (\ref{13}), (\ref{13c+}) ((\ref{13c-})) define the mapping
\be
\phi^{+}_{1}: {\bf G_{+}}\to {\bf C_{1}} \label{13m+}
\en
\be
(\phi^{-}_{1}: {\bf G_{-}}\to {\bf C_{1}}) \label{13m-}
\en

 For the $N=2$ SWZNW model on the group ${\bf G}$ we obtain from
(\ref{13}) the decompositions for the superfield (\ref{4}) (which takes
values in ${\bf C_{1}}$)
\be
G(x_{+},x_{-})= G_{+}(x_{+},x_{-})G^{-1}_{-}(x_{+},x_{-})=
                {\tld G}_{-}(x_{+},x_{-}){\tld G}^{-1}_{+}(x_{+},x_{-})
\label{14}
\en

 To generalize (\ref{13}), (\ref{13u}) one have to consider the
set $W$ (which we shall assume in the following to be discret and
finite set) of classes ${\bf G_{+}}\backslash {\bf G^{\Bbb C}}/ {\bf G_{-}}$
and pick up a representative $w$ for each class $[w]\in W$.
It gives us the stratification
of ${\bf G^{\Bbb C}}$ ~\cite{AlMal}:
\be
{\bf G^{\Bbb C}}= \bigcup_{[w]\in W} {\bf G_{+}}w{\bf G_{-}}=
         \bigcup_{[w]\in W} {\bf G_{w}}  \label{17+}
\en
There is the second stratification:
\be
{\bf G^{\Bbb C}}= \bigcup_{[w]\in W} {\bf G_{-}}w{\bf G_{+}}=
         \bigcup_{[w]\in W} {\bf G^{w}}  \label{17-}
\en
We shall assume, in the following, that the representatives $w$
have picked up to satisfy the unitarity condition:
\be
\tau (w)=w^{-1}                        \label{17w}
\en
It allows us to generalize (\ref{13}), (\ref{13u}) as follows
\be
g= wg_{+}g^{-1}_{-}= w{\tld g}_{-}{\tld g}^{-1}_{+},    \label{18t}
\en
where
\be
g_{+}\in {\bf G^{w}_{+}},\
{\tld g}_{-}\in {\bf G^{w}_{-}} \label{037}
\en
and
\be
{\bf G^{w}_{+}}= {\bf G_{+}}\cap w^{-1}{\bf G_{+}}w, \
{\bf G^{w}_{-}}= {\bf G_{-}}\cap w^{-1}{\bf G_{-}}w.
\label{038}
\en

 It is clear that there exists also an appropriate generalization
of (\ref{14}) for the decompositions (\ref{18t}).

 In order to the element $g$ belongs to the real group ${\bf G}$
the elements $g_{\pm}, {\tld g}_{\pm}$ from (\ref{18t})
should satisfy (\ref{13u}).
Thus the formulas (\ref{18t}, \ref{037}), (\ref{13c+}) ((\ref{13c-}))
define the mapping
\be
\phi^{+}_{w}: {\bf G^{w}_{+}}\to {\bf C_{w}}\equiv {\bf G_{w}}\cap {\bf G}
\label{037m}
\en
\be
(\phi^{-}_{w}: {\bf G^{w}_{-}}\to {\bf C_{w}}\equiv {\bf G_{w}}\cap {\bf G}).
\label{037m-}
\en

To formulate the main result of the paper ~\cite{P2}
one needs to introduce some notations.
Let
\be
\lbrc R_{i}, i=1,...,d\rbrc, \label{Rb+}
\en
be the basis in the Lie subalgebra ${\bf g_{+}}$ and
\be
\lbrc R^{i}, i=1,...,d\rbrc, \label{Rb-}
\en
be the basis in the Lie subalgebra ${\bf g_{-}}$
so that (\ref{Rb+}, \ref{Rb-}) constitute the orthonormal basis
in ${\bf g^{\Bbb C}}$:
\be
<R^{i}, R_{j}>=\delta^{i}_{j}. \label{Rbn}
\en
We identify the Lie algebra ${\bf g^{\Bbb C}}$ with the
space of complex left invariant vector fields on the group ${\bf G}$.
To each decomposition (\ref{14}) or its generalization
for the mappings into others strats ${\bf C_{w}}$ we introduce
the superfields
\ber
\rh^{+}=G^{-1}_{+}DG_{+}=\rh^{i}R_{i}, \
\rh^{-}=G^{-1}_{-}DG_{-}=\rh_{i}R^{i}, \nmb
{\tld \rh}^{+}={\tld G}^{-1}_{+}D{\tld G}_{+}={\tld \rh}^{i}R_{i}, \
{\tld \rh}^{-}={\tld G}^{-1}_{-}D{\tld G}_{-}={\tld \rh}_{i}R^{i}.
\label{not1}
\enr
These superfields correspond to the left invariant 1-forms on
${\bf G_{\pm}}$
\be
r^{\pm}=g^{-1}_{\pm}dg_{\pm}, \
{\tld r}^{\pm}={\tld g}^{-1}_{\pm}d{\tld g}_{\pm}.
\label{not2}
\en

 In ~\cite{P2} the following statements was proved:
\begin{itemize}
\item
the mappings (\ref{037m})
are holomorphic and define the natural action of the complex group
${\bf G_{+}}$  on ${\bf G}$ generated by the holomorphic vector
fields $\lbrc S_{i}, i=1,...,d\rbrc$;
the set $W$ parametrizes ${\bf G_{+}}$-orbits ${\bf C_{w}}$.

\item
The Lagrangian $\Lm $ of the model is given by
\be
\Lm={\im \ov 2}\Om_{cb}J^{c}_{a}\rh^{a}_{+}\rh^{b}_{-}, \label{LGR}
\en
where $\Om_{cb}$ is Semenov-Tian-Shansky symplectic form
on ${\bf G}$ ~\cite{SemTian}, $J^{c}_{a}$ is left invariant complex
structure on ${\bf G}$ and we have used common notation
$\rh^{a}, a=1,...,2d$ for the 1-forms $\rh^{i}, \bar{\rh}^{i}$.
Remark that stratifications (\ref{17+}), (\ref{17-}) code
degegnerations of $\Om$ ~\cite{AlMal} thus the
formula (\ref{LGR}) is true within each strat ${\bf C_{w}}$.

\item
$({\bf G},J)$-SWZNW model admits PL symmetry
with respect to the holomorphic ${\bf G_{+}}$-action,
i.e. the following conditions are satisfied on the extremals
\ber
L_{S_{i}}\Lm= f^{jk}_{i}(A_{+})_{j}(A_{-})_{k} \nmb
L_{\bar{S}_{i}}\Lm= \bar{f}^{jk}_{i}(A_{+})_{\bar{j}}(A_{-})_{\bar{k}},
\label{PLS}
\enr
where $L_{S_{i}}, L_{\bar{S}_{i}}$ mean the Lie derivatives along
the vector fields $S_{i}, \bar{S}_{i}$ ($\bar{S}_{i}$ are complex
conjugated to $S_{i}$),
$f^{ik}_{j}$ are the structure constants of the Lie algebra
${\bf g_{-}}$ ($\bar{f}^{jk}_{i}$ are complex conjugated to
$f^{ik}_{j}$) and the Noether currents $A_{i}, A_{\bar{i}}$
are given by
\be
(A_{-})_{i}=(\rh_{-})_{i}, \
(A_{+})_{i}=\im (J\rh_{+})_{i}, \
(A_{\pm})_{\bar{i}}= ({\bar A}_{\pm})_{i}.
\label{NC}
\en

\end{itemize}

 The equations (\ref{PLS}) are equivalent
to zero curvature equations for the $F_{+-}$-component of the super
stress tensor $F_{MN}$ ~\cite{KlimS1, P2} 
\ber
(F_{+-})_{i}\equiv D_{+}(A_{-})_{i}+D_{-}(A_{+})_{i}-
              f^{nm}_{i}(A_{+})_{n}(A_{-})_{m}=0  \nmb
(F_{+-})_{\bar{i}}\equiv D_{+}(A_{-})_{\bar{i}}+D_{-}(A_{+})_{\bar{i}}-
              \bar{f}^{nm}_{i}(A_{+})_{\bar{n}}(A_{-})_{\bar{m}}=0
\label{ZC1}
\enr
Using the standard arguments of the super Lax construction ~\cite{EvHol}
one can show that from (\ref{ZC1}) it follows that the connection is flat
\be
F_{MN}=0,\ M, N= (+, -, +, -).  \label{ZC}
\en

  With the appropriate modifications the above statements
are true also for the mappings (\ref{037m-}) and ${\bf G_{-}}$-action
on ${\bf G}$. Due to this observation PL self-duality $({\bf G},J)$-
SWZNW models was proved in ~\cite{P2}. 

\vskip 10pt
\centerline{\bf 2. PL T-self-duality in the $N=2$ SWZNW models.}


 In this section we consider PL T-duality in  $({\bf G},J)$-SWZNW models
and obtain transformation formulas of the super-Kac-Moody currents  
following to ~\cite{KlimS1, P2}.


 Due to the equation (\ref{ZC}) we may associate to
each extremal surface $G_{+}(x_{+}, x_{-}, \Tta_{+}, \Tta_{-})\in {\bf G_{+}}$,
a mapping ("Noether charge") $V_{-}(x_{+}, x_{-}, \Tta_{+}, \Tta_{-})$
from the super world-sheet into the group ${\bf G_{-}}$ such that
\be
(A_{\pm})_{i}=-(D_{\pm}V_{-}V^{-1}_{-})_{i}.
\label{41}
\en
Now we build the following surface in the
double ${\bf G^{\Bbb C}}$:
\ber
F(x_{+}, x_{-}, \Tta_{+}, \Tta_{-})=
G_{+}(x_{+}, x_{-}, \Tta_{+}, \Tta_{-})
V_{-}(x_{+}, x_{-}, \Tta_{+}, \Tta_{-}).
\label{42}
\enr
In view of (\ref{NC},\ref{not1}) it is natural to represent $V_{-}$ as the
product
\be
V_{-}=G^{-1}_{-}H^{-1}_{-}
\label{42p}
\en
, where $G_{-}$ is determined
from (\ref{13c+}) and $H_{-}$ satisfies the equation
\be
D_{-}H_{-}=0. \label{42der}
\en
Therefore the surface (\ref{42}) can be rewritten in the form
\be
F(x_{\pm}, \Tta_{\pm})=G(x_{\pm}, \Tta_{\pm})H^{-1}_{-}(x_{+}, \Tta_{-}),
\label{42.1}
\en
where $G(x_{\pm}, \Tta_{\pm})\in {\bf G}$ is the solution of
$({\bf G})$-SWZNW model. Now we represent the ${\bf G_{-}}$-valued
field $H_{-}$ in terms of conservation currents $I_{+}\equiv G^{-1}D_{+}G$
of the model. To do it let's denote by $\xi $ the canonic
${\bf g^{\Bbb C}}$-valued left invariant 1-form on the group ${\bf G}$.
It is obvious that
\ber
\xi= \xi ^{j}R_{j}+ \xi _{j}R^{j}, \nmb
J\xi^{k}=-\im \xi ^{k}, \ J\xi_{k}= \im \xi _{k}.
\label{1Form}
\enr
 Using the first decomposition from (\ref{13}) we get
\be
\xi= g_{-}r^{+}g^{-1}_{-}-l^{-}. \label{1Forr}
\en
Let's introduce the matrices
\ber
g_{-}R_{i}g^{-1}_{-}=M_{ij}R^{j}+N^{j}_{i}R_{j},\nmb
g_{+}R^{i}g^{-1}_{+}=P^{ij}R_{j}+Q^{i}_{j}R^{j},\nmb
g_{-}R^{i}g^{-1}_{-}=(N^{\ast})^{i}_{j}R^{j},\nmb
g_{+}R_{i}g^{-1}_{+}=(Q^{\ast})^{j}_{i}R_{j}.
\label{Matr}
\enr
Using these matrices and (\ref{Rbn}) we can express the
1-forms $r_{i}$, in terms of $\xi ^{i}$:
\be
r_{i}=((N^{\ast})^{-1})^{j}_{i}(-\xi _{j}+M_{nj}(N^{-1})^{n}_{k}\xi ^{k}).
\label{Jr1}
\en
Then
\be
\im Jr_{i}=((N^{\ast})^{-1})^{j}_{i}(\xi _{j}+M_{nj}(N^{-1})^{n}_{k}\xi ^{k}).
\label{Jr2}
\en
From the other hand in view of (\ref{42p}) we shall get
\be
(A_{+})_{i}= (\rh _{+})_{i}+ (G^{-1}_{-}H^{-1}_{-}D_{+}H_{-}G_{-})_{i}.
\label{Jr3}
\en
Comparing (\ref{Jr2}) and (\ref{Jr3}) and taking into account (\ref{NC}),
(\ref{Matr}) we conclude  
\be
H^{-1}_{-}D_{+}H_{-}=2(I_{+})^{-},
\label{Jr4}
\en
where $(I_{+})^{-}$ is ${\bf g_{-}}$-projection of $I_{+}$. 
Though the formula (\ref{Jr4}) is obtained for
the mappings into the strat ${\bf C_{1}}$ we claim that it remains to be
true for the mappings into another strats, so that (\ref{42.1}) 
is correct on the super world-sheet everywhere. 

 The solution and the "Noether charge"
of the dual $\sgm $-model are given by "dual"
parametrization of the surface (\ref{42}) ~\cite{KlimS1}
\ber
F(x_{+}, x_{-}, \Tta_{+}, \Tta_{-})=
\brv{G}_{-}(x_{+}, x_{-}, \Tta_{+}, \Tta_{-})
\brv{V}_{+}(x_{+}, x_{-}, \Tta_{+}, \Tta_{-}),
\label{42d}
\enr
where $\brv{G}_{-}(x_{+}, x_{-}, \Tta_{+}, \Tta_{-})\in {\bf G_{-}}$
and $\brv{V}_{+}(x_{+}, x_{-}, \Tta_{+}, \Tta_{-})\in {\bf G_{+}}$.
Thus in the dual $\sgm$-model Drinfeld's dual group to
the group ${\bf G_{+}}$ should acts, i.e. it should be
a $\sgm$-model on the orbits of the group ${\bf G_{-}}$
 and with respect to this action
the dual to (\ref{PLS}) PL symmetry conditions should be satisfied:
\ber
L_{S^{i}}\brv{\Lm}= f_{jk}^{i}(\brv{A}_{+})^{j}(\brv{A}_{-})^{k}, \nmb
L_{\bar{S}^{i}}\brv{\Lm}=
\bar{f}_{jk}^{i}(\brv{A}_{+})^{\bar{j}}(\brv{A}_{-})^{\bar{k}},
\label{PLSd}
\enr
where $\lbrc S^{i}, \bar{S}^{i}, i=1,...,d\rbrc $ are the vector fields
which generate the ${\bf G_{-}}$-action,
$\brv{\Lm}, \brv{A}_{\pm}^{j}, \brv{A}_{\pm}^{\bar{j}}$
are the Lagrangian and the Noether currents in the dual $\sgm$-model.
 It was argued in ~\cite{P2} $({\bf G},J)$-SWZNW models
are PL self-dual models and (\ref{42d}) can be rewritten as follows
\be
F(x_{\pm}, \Tta_{\pm})=
\brv{G}(x_{\pm}, \Tta_{\pm})H^{-1}_{+}(x_{+}, \Tta_{-})
,\label{48}
\en
where $\brv{G}(x_{\pm}, \Tta_{\pm})$ is the dual
solution of ${\bf G}$-SWZNW model. Similar to the equation
(\ref{Jr4}) we can obtain
\be
H^{-1}_{+}D_{+}H_{+}=2(\brv{I}_{+})^{+},
\label{Jr5}
\en
where $(\brv{I}_{+})^{+}$ is ${\bf g_{+}}$-projection of
$\brv{I}_{+}\equiv \brv{G}^{-1}D_{+}\brv{G}$. 

 From (\ref{Jr4},\ref{Jr5}) we conclude that under PL T-duality
\be
t: G(x_{\pm}, \Tta_{\pm}) \to \brv{G}(x_{\pm}, \Tta_{\pm})
\label{tmap}
\en 
the conserved current $I_{+}$ transforms as  
\be
t: (I_{+})^{-}\to (\brv{I}_{+})^{+}, \
   (I_{+})^{+}\to (\brv{I}_{+})^{-}.
\label{Imap}
\en
Moreover as it follows from (\ref{42.1},\ref{48})
the conserved currents $I_{-}\equiv D_{-}GG^{-1}$ map
under PL T-duality identicaly:
\be
t: (I_{-})^{\pm}\to (\brv{I}_{-})^{\pm}.
\label{Imap-}
\en


\vskip 10pt
\centerline{\bf3. PL T-duality and Mirror symmetry.}

 In this section we consider the behaviour of the $N=2$ super-Virasoro
algebra generators under PL T-duality.
 Let's consider at first the transformation low for the components
of the supercurrent $I_{+}$. On the extremals of ${\bf G}$-SWZNW
model we have the following expansion of $I_{+}$
\be
I_{+}=\im g^{-1}\psi _{+}
-\im \Tta _{-}(g^{-1}\d _{+}g +\im g^{-1}\psi _{+}g^{-1}\psi_{+}).
\label{51}
\en 
We introduce the notations:
\ber
\phi _{+}=g^{-1}\psi _{+},\
q_{+}=g^{-1}\d_{+}g,      \nmb
j_{+}=q_{+}+{\im \ov 2}\lbrc \phi_{+},\phi_{+}\rbrc.
\label{52}
\enr
Then
\be
I_{+}=\im (\phi_{+}-\Tta_{-}j_{+}).
\label{53}
\en
Remark that fields $\phi_{+}$ introduced in (\ref{52})
are the free fermions:
\be
\d_{-}\phi_{+}=0
\label{54}
\en
and with the conserved currents $j_{+}$ they generate (left-moving) 
N=1 super-Kac-Moody algebra of the model ~\cite{swzw, KcT}.
It is easy to obtain from (\ref{Imap}) PL T-duality transformation low
for the currents (\ref{52})
\be
t: \phi^{\pm}_{+}\to \brv{\phi}^{\mp}_{+},\
j^{\pm}_{+}\to \brv{j}^{\mp}_{+}.
\label{55}
\en
 The generator of N=1 superconformal symmetry of the model is given by
(we shall omit in the following the world sheet indeces ${\pm}$
having in mind that we are working in the left-moving sector) 
\ber
\Gm =<DI,I>+{1 \ov 3}<\lbrc I,I\rbrc,I>= \nmb
<q,\phi>+{\im \ov 6}<\lbrc \phi,\phi \rbrc,\phi>-
\Tta_{-}(<q,q>-\im <\d\phi,\phi>)\eq
\Sgm +\Tta_{-}2T.
\label{56}
\enr
It can be represented as the sum
\ber
\Gm= \Gm^{+}+\Gm^{-}, \nmb
\Gm^{\pm}=<q^{\pm},\phi^{\mp}>+
{\im \ov 2}<\lbrc \phi^{\pm},\phi^{\pm} \rbrc,\phi^{\mp}>-
\Tta_{-}(<q^{\pm},q^{\mp}>-\im <\d\phi^{\pm},\phi^{\mp}>).
\label{57}
\enr
Thus we recognize spin 3/2- currents
\be
\Sgm^{\pm}=<q^{\pm},\phi^{\mp}>+
{\im \ov 2}<\lbrc \phi^{\pm},\phi^{\pm} \rbrc,\phi^{\mp}>,
\label{58.1}
\en
stress-energy tensor
\be
T=-{1\ov 2}(<q,q>-\im<\d \phi,\phi>)
\label{58.2}
\en
and $U(1)$ current
\be
K=<\phi^{+},\phi^{-}>
\label{58.3}
\en
of the $N=2$ super-Virasoro algebra so that
\be
\Gm^{\pm}=\Sgm^{\pm}+\Tta_{-}(T\pm {\im \ov 2} \d K).
\label{59}
\en
Using (\ref{52}) we rewrite (\ref{59}) in more convenient form
\be
\Gm^{\pm}=<j^{\pm},\phi^{\mp}>-
\im<\lbrc \phi^{\mp}, \phi^{\mp}\rbrc,\phi^{\pm}>-
\Tta_{-}({1\ov 2}(<j,j>-<\lbrc \phi,\phi\rbrc,j>)-
\im<\d \phi^{\pm},\phi^{\mp}>).
\label{60}
\en
From (\ref{55}) we obtain PL T-duality mapping for $\Gm^{\pm}$:
\be
t: \Gm^{\pm}\to \brv{\Gm}^{\mp},
\label{61}
\en
or in components
\be
t: \Sgm^{\pm}\to \brv{\Sgm}^{\mp}, \
T\pm \im \d K \to \brv{T}\mp \im \d \brv{K}.
\label{62}
\en
Thus PL T-duality in the $N=2$ SWZNW models is a mirror duality.
With the result of ~\cite{P2} that PL T-duality is some special
super-Kac-Moody transformation it gives us more rigorous background
for the conjecture proposed in ~\cite{GiWi}.
 
 Now we discuss Kazama-Suzuki models generalization of PL T-duality.
Kazama and Suzuki have studied  ~\cite{KaSu} under what conditions an N=1
superconformal coset model can have an extra supersymmetry,
to give rise to an $N=2$ superconformal model. 
Their conclusion can be formulated in terms of Manin triples:
an $N=2$ superconformal coset model is defined by a Manin triple 
$({\bf g},{\bf g_{+}},{\bf g_{-}})$ and Manin subtriple
$({\bf h},{\bf h_{+}},{\bf h_{-}})$, ${\bf h}\in {\bf g},
{\bf h_{\pm}}\in {\bf g_{\pm}}$ so that the subspaces
${\bf l_{\pm}\eq g_{\pm}}/ {\bf h_{\pm}}$ are the Lie
algebras ~\cite{HulS}.
The spin-3/2 currents, stress-energy tensor and $U(1)$current
of the ${\bf G/H}$-coset model are given respectively by
\ber
\Sgm_{c}^{\pm}=<q^{\pm}_{l},\phi^{\mp}_{l}>+
{\im \ov 2}<\lbrc \phi^{\pm}_{l},\phi^{\pm}_{l} \rbrc,\phi^{\mp}_{l}>, \nmb
T_{c}=-{1\ov 2}(<q_{l},q_{l}>-\im<\d \phi_{l},\phi_{l}>), \nmb
K_{c}=<\phi^{+}_{l},\phi^{-}_{l}>,
\label{58.KS}
\enr
where the subscript $l$ means the projection on the subspace
${\bf l=g/h}$.

 Due to the denominator of the coset is
$N=2$ SWZNW model on the subgroup ${\bf H\in G}$ PL T-duality
maps the currents of this model by the formulas
(\ref{55},\ref{62}). Thus if we define PL T-duality transformation
in ${\bf G/H}$-Kazama Suzuki coset model as an appropriate 
projection of the transformation in ${\bf G}$-SWZNW model we obtain,
taking into account (\ref{58.KS}), the formulas (\ref{62}) as the
transformation low of the $N=2$ super-Virasoro algebra generators of the coset.
Though in order to establish accurately PL T-duality in Kazama-Suzuki
models an individual investigation should be done, we can conclude from
these arguments that PL T-duality in the Kazama-Suzuki models is 
a mirror duality also. 

 It is pertinent to note that in contrast to the $N=2$ SWZNW models
the PL T-duality in Kazama-Suzuki models can be interpreted as a 
mirror symmetry between the K$\ddot{a}$hler manifolds. Indeed as it 
follows from the results
of the paper ~\cite{SevP} the nessesary condition for such interpretation
is the realization of the (2,2) $\sgm$-model solely by chiral or solely
by twisted chiral superfields. In this case 
the left-moving and right-moving
spin-3/2 currents of the model are constructed with help of
only one complex structure on the manifold so that we can 
associate the chiral-chiral and chiral-antichiral rings of the model 
with the De Ram and Dol'bo cohomology of a manifold ~\cite{MS}. 
Thus PL T-duality in the Kazama-Suzuki model
can be related with the mirror symmetry of the K$\ddot{a}$hler
manifolds if all primary fields of this model are only chiral
or only twisted chiral fields. An intresting question
which arises in this context is is there an analog of the ${\bf G_{+}}$-
holomorphic action and PL symmetries in Calabi-Yau manifolds.

 The discussion in this note was puerly classical.
The natural and most important question what is the quantum picture of 
the PL T-duality in the $N=2$ superconformal theories. Because the
Poisson-Lie groups is nothing but a classical limit of the quantum groups 
~\cite{Drinf1} there appears an intriguing possibility of a relevance 
of quantum groups in the T-duality. In particular it would appear reasonable
that mirror symmetry closely related with quantum group duality principle
~\cite{SemTian1}.

\vskip 10pt
\centerline{\bf ACKNOWLEDGEMENTS}
\frenchspacing

 I am very grateful to A. Alekseev for helpful discussions
and careful reading of the manuscript of the paper ~\cite{P2}.
This work was supported in part by grants
INTAS-95-IN-RU-690, CRDF RP1-277, RFBR 96-02-16507.

\vfill
\end{document}